# Striped magnetization plateau and chirality-reversible anomalous Hall effect in a magnetic kagome metal

Erjian Cheng[1,#,*], Ning Mao[1,#], Xiaotian Yang[2,#], Boqing Song[3,#], Rui Lou[4,5,6,#,*], Tianping Ying[3], Simin Nie[7], Alexander Fedorov[4,5,6], François Bertran[8], Pengfei Ding[9], Oleksandr Suvorov[4,10], Shu Zhang[11], Susmita Changdar[4], Walter Schnelle[1], Ralf Koban[1], Changjiang Yi[1], Ulrich Burkhardt[1], Bernd Büchner[4,12], Shancai Wang[9], Yang Zhang[13,14*], Wenbo Wang[2,*] and Claudia Felser[1,*]

[1] *Max Planck Institute for Chemical Physics of Solids, 01187 Dresden, Germany*

[2] *School of Physical Science and Technology, ShanghaiTech Laboratory for Topological Physics, ShanghaiTech University, 201210 Shanghai, China*

[3] *Institute of Physics and University of Chinese Academy of Sciences, Chinese Academy of Sciences, 100190 Beijing, China*

[4] *Leibniz Institute for Solid State and Materials Research, IFW Dresden, 01069 Dresden, Germany*

[5] *Helmholtz-Zentrum Berlin für Materialien und Energie, Albert-Einstein-Straße 15, 12489 Berlin, Germany*

[6] *Joint Laboratory "Functional Quantum Materials" at BESSY II, 12489 Berlin, Germany*

[7] *Department of Mechanical Engineering, Stanford University, 94305 Stanford, California, USA*

[8] *Synchrotron SOLEIL, L'Orme des Merisiers, Départementale 128, 91190 Saint-Aubin, France*

[9] *Department of Physics, Key Laboratory of Quantum State Construction and Manipulation (Ministry of Education), and Beijing Key Laboratory of Opto-electronic Functional Materials & Micronano Devices, Renmin University of China, 100872 Beijing, China*

[10] *Kyiv Academic University, 03142 Kyiv, Ukraine*

[11] *Max Planck Institute for the Physics of Complex Systems, 01187 Dresden, Germany*

[12] *Institute for Solid State and Materials Physics, TU Dresden, 01062 Dresden, Germany*

[13] *Department of Physics and Astronomy, University of Tennessee, Knoxville, 37996 Tennessee, USA*

[14] *Min H. Kao Department of Electrical Engineering and Computer Science, University of Tennessee, Knoxville, 37996 Tennessee, USA*

**Abstract**

Kagome materials with magnetic frustration in two-dimensional networks are known for their exotic properties, such as the anomalous Hall effect (AHE) with non-collinear spin textures. However, the effects of one-dimensional (1D) spin chains within these networks are less understood. Here, we report a distinctive AHE in the bilayer-distorted kagome material GdTi$_3$Bi$_4$, featuring 1D Gd zigzag spin chains, a one-third magnetization plateau, and two successive metamagnetic transitions. At these metamagnetic transitions, Hall resistivity shows abrupt jumps linked to the formation of stripe domain walls, while within the plateau, the absence of detectable domain walls suggests possible presence of skyrmion phase. Reducing the sample size to a few microns reveals additional Hall resistivity spikes, indicating domain wall skew scattering contributions. Magnetic atomistic spin dynamics simulations reveal that the magnetic textures at these transitions have reverse chirality, explaining the evolution of AHE and domain walls with fields. These results underscore the potential of magnetic and crystal symmetry interplay, and magnetic field-engineered spin chirality, for controlling domain walls and tuning transverse properties, advancing spintronic applications.

**Introduction**

Kagome-lattice materials with a two-dimensional (2D) network of corner-sharing triangles have garnered significant interest due to their alluring phenomena and emergent states of matter, such as flat bands, van Hove singularities (vHSs), Dirac fermions, unconventional superconductivity, density wave orders, and electronic nematicity [1–8]. In magnetic cases, different from ordinary Hall effect due to the Lorentz force deflecting the motion paths of charge carriers, the inclusion of magnetism with time-reversal-symmetry breaking could give rise to spontaneous AHE [9,10]. Within a non-collinear spin configuration, conduction electrons passing through this spin texture acquire a quantum-mechanical Berry phase, resulting in a local fictitious magnetic field proportional to the scalar spin chirality, $\chi_{ijk} = \boldsymbol{S}_i \cdot (\boldsymbol{S}_j \times \boldsymbol{S}_k)$, where $\boldsymbol{S}_{i,j,k}$ are spins, leading to an AHE or so-called topological Hall effect (THE) whose amplitude scales with this fictitious field rather than magnetization [11,12]. This has been usually observed in noncentrosymmetric magnets stabilized by the Dzyaloshinskii–Moriya interaction (DMI), such as MnSi, MnGe, and FeGe [13–18]. Alternatively, THE can arise from magnetic frustration in materials without DMI, like Gd$_2$PdSi$_3$ with a triangular lattice, Pr$_2$Ir$_2$O$_7$ [20] and Nd$_2$Mo$_2$O$_7$ [21] with pyrochlore structures, and Mn$_3$Sn [22], Mn$_3$Ga [23], Fe$_3$Sn$_2$ [24] and $R$Mn$_6$Sn$_6$ ($R$ = rare earth) [25–29]. These frustrated magnets typically have quasi-2D or three-dimensional (3D) magnetic networks, while systems containing (quasi-)1D spin chains are less explored. Spin chains, which exhibit unique properties such as strong quantum fluctuations, the breakdown of Landau's Fermi liquid theory, and exact solutions in some models, present a compelling area of study [30,31]. Thus, exploring the physical properties of materials with spin chains constrained by high crystal symmetries in kagome systems is highly appealing.

$Ln$Ti$_3$Bi$_4$ ($Ln$ = rare-earth elements) family with quasi-1D spin chains running along the $a$ axis have recently emerged as titanium-based bilayer kagome metals [32–38]. These materials exhibit various and tunable magnetic properties through rare-earth engineering [32–38]. $Ln$Ti$_3$Bi$_4$ crystallizes in an orthorhombic structure within the $Fmmm$ (No. 69) space group with inversion symmetry. This structure consists of alternating layers of Ti$_3$Bi$_4$, $Ln$Bi$_2$, and Bi along the $c$ axis (Figs. 1a and 1b). Unlike the $D_{6h}$ symmetry observed in kagome metals $AM_3$Sb$_5$ (where $A$ = K, Rb, Cs, $M$ = V, Ti) [1–3, 39–42], the quasi-1D nature of the $Ln$ zigzag chains along the $a$ axis leads to an orthorhombic structure in the $Ln$Bi$_2$ layer, resulting in reduced crystalline symmetry ($D_{2h}$). Among $Ln$Ti$_3$Bi$_4$, GdTi$_3$Bi$_4$ orders as an antiferromagnetic metal with two successive magnetic transitions at $T_N^{'} \sim 14.1$ K and $T_N \sim 13.3$ K, respectively (Fig. 1d and Supplementary Fig. 2a) [32]. More interestingly, when a magnetic field is applied parallel to the $c$ axis, the magnetization exhibits nearly one-third of its saturation value (Fig. 1e), suggesting magnetic frustration in GdTi$_3$Bi$_4$ [42–45]. We conducted angle-resolved photoemission spectroscopy (ARPES) to characterize the electronic band structure of GdTi$_3$Bi$_4$, revealing double flat bands, and multiple vHSs. Intriguingly, bulk band splittings in both magnetic and paramagnetic states have been unveiled, in sharp contrast to other $Ln$Ti$_3$Bi$_4$ siblings [32–38]. Electrical transport, magnetic force microscopy (MFM) measurements, and theoretical simulations reveal a unique AHE possibly associated with a skyrmion phase, and stripe domain walls with reverse chirality at the two successive metamagnetic transitions (entering and exiting magnetization plateau).

## Results and discussion

**Electronic band structure of GdTi$_3$Bi$_4$**

Figure 1c displays the 3D orthorhombic Brillouin zone (BZ) together with the pseudo-hexagonal projected 2D BZ. The reduced (two-fold) rotational symmetry is reflected in the pseudo-hexagonal projection featuring two equivalent $\bar{M}$ and $\bar{K}'$ points, respectively. The ARPES constant energy maps at various binding energies are presented in Fig. 2f. The overall Fermi surface (FS) exhibits a two-fold symmetry, akin to other $Ln$Ti$_3$Bi$_4$ compounds [32–38], with respect to the $\bar{\Gamma}-\bar{M}$ or $\bar{\Gamma}-\bar{K}'$ directions. The revealed breaking of the six-fold rotational symmetry, which stems from the Ti-based kagome sublattice, is necessitated by the quasi-1D nature of the Gd chains. Figures 2g–2i show the near-$E_F$ band dispersions recorded along the $\bar{\Gamma}-\bar{K}-\bar{M}'$ lines under different photon energies. Two electron-like bands and multiple linear dispersions are observed around $\bar{\Gamma}$ and $\bar{K}/\bar{M}'$ points, respectively. These spectra match well with our nonmagnetic DFT calculations in Fig. 2j. In addition, as marked out by the grey shades in Figs. 1g and 1h, one can identify two nearly dispersionless bands at around -0.4 and -0.7 eV running over a wide range of momentum. They are in good agreement with the kagome flat bands (FBs) in DFT calculations (Fig. 1j), signifying their phase-destructive-FB origin [46] (see Supplementary Fig. 3 and Note 3 for the corroboration of another band characteristic of kagome lattice, the van Hove singularities).

We then study the magnetic impact on the electronic band structure of GdTi$_3$Bi$_4$. Intriguingly, despite the above measurements being conducted in the paramagnetic phase of GdTi$_3$Bi$_4$, upon a closer scrutiny of the outer electron band ($\beta$) at $\bar{\Gamma}$ point (Figs. 1g and 1h), one could identify some faint signs of the band splitting. In Figs. 1k and 1l, the high-resolution cut along the $\bar{\Gamma}-\bar{K}-\bar{M}'$ direction obtained by 83-eV photons has been displayed, where the $\beta$ band is more clearly visualized. As depicted in Fig. 1l, one reveals a noticeable Zeeman-like splitting of the $\beta$ band, as evidenced by the "kink"-like feature (indicated by the green arrows) and the momentum distribution curve (MDC) taken at $E_F$ (red solid curve in the inset). Similar splitting is also observed on the $\alpha$ band around the second BZ center (Figs. 1n and 1o) and shows no observable evolution after cooling to below $T_N$ (Figs. 1p and 1q). Recently, similar Zeeman-like band splittings (on the two electron-like bands around the BZ center) have been reported in ferromagnets SmTi$_3$Bi$_4$ [34] and NdTi$_3$Bi$_4$ [37], but only within the ferromagnetic phases. In our case of GdTi$_3$Bi$_4$, the band splittings are quite unusual because both inversion symmetry and time-reversal symmetry are preserved in the paramagnetic phase. Moreover, as discussed in Supplementary Note 4, the magnetic coupling strength in the topmost Gd layer is comparable with the bulk, ruling out the scenario of the surface enhancement of static magnetic order [47,48]; the overall good correspondence between experiments and DFT (Fig. 1j) further excludes the possibility that the band splitting results from the $k_z$ broadening effect in photoemission process [49].

A comparable instance of unusual band splitting has been observed in the antiferromagnet EuCd$_2$As$_2$, which is attributed to ferromagnetic fluctuations within the paramagnetic region [50]. In the case of GdTi$_3$Bi$_4$, the positivity of Curie-Weiss temperature unambiguously suggests the existence of ferromagnetic coupling (Supplementary Fig. 2b), reflecting predominantly the bulk magnetic property of the band splitting. Consistently, by introducing a ferromagnetic order into the Gd sites, the DFT calculations well reproduce the split $\alpha$ and $\beta$ bands (Fig. 1m). More discussion on the magnetic ground state and the magnetic impact on electronic structures can be found in Supplementary Note 5. The revealed ferromagnetic fluctuations in the electronic band structure suggest that the complex magnetic texture might be at play in transport.

**Anomalous transverse transport and stripe domain walls in GdTi$_3$Bi$_4$**

Figure 2a shows the transverse Hall resistivity profiles with the current applied along the *a* axis and the magnetic field along the *c* axis. At 2 K, as the magnetic field increases from 0 to 9 T, the Hall resistivity exhibits a jump at $H_1 \sim 1.71$ T and then decreases sharply at $H_2 \sim 3.43$ T, corresponding to the two metamagnetic transitions (Figs. 2c and 2d). Beyond $H_2$, where the magnetization increases more gradually, the Hall resistivity displays an additional hump. When the magnetic field surpasses the polarized field ($H_3 \sim 4.5$ T), the Hall resistivity exhibits non-linear behavior, suggesting the presence of multiple band features in GdTi$_3$Bi$_4$, as observed in ARPES experiments, and indicating a minimal contribution from the spontaneous anomalous Hall effect. With increasing temperatures, the Hall peaks are gradually weakened, and the profile at 16 K slightly larger than the ordering temperature shows no distinct anomaly. Figure 2b shows the MR with the same measured configuration, which displays similar behavior as Hall resistivity. GdTi$_3$Bi$_4$ orders at 14.1 K, but the MR remains negative up to 30 K, suggesting a field-induced suppression of scattering from local moments. This observation also points to strong magnetic fluctuations in the paramagnetic state, aligning with the band splitting seen in ARPES measurements.

For Hall resistivity, the empirical relation: $\rho_{yx} = \rho_{yx}^O + \rho_{yx}^A + \rho_{yx}^T$ has been usually applied to separate the different contributions to $\rho_{yx}$. Here, $\rho_{yx}^O$, $\rho_{yx}^A$ and $\rho_{yx}^T$ are the ordinary, anomalous Hall, and topological Hall contribution, respectively. For GdTi$_3$Bi$_4$, the small contribution of spontaneous anomalous Hall effects—evidenced by the nonlinear behavior of Hall resistivity in saturated fields—and the presence of complex Fermi surfaces near the Fermi level make it challenging to distinguish these three contributions through fitting. Therefore, to determine the anomalous Hall contribution associated with magnetic textures, we use the data at 16 K as a reference for subtraction, assuming that the carrier density and mobility remain constant. In this work, we continue to use the nominal $\rho_{yx}^A$ for discussion, as we did not separate the spontaneous anomalous Hall and topological Hall contributions. The obtained data is shown in Fig. 2e. For better comparison, the magnetic field *vs.* temperature phase diagram for magnetization has been plotted, which is derived from the temperature dependence of magnetization in increasing magnetic fields during zero-field-cooling process (Supplementary Fig. 2c), as illustrated in Fig. 2f. The successive magnetic transitions at $T_N^{'}$ and $T_N$ display different magnetic field dependence with decreasing temperature, and the derivative of normalized magnetic susceptibility associated with two successive metamagnetic transitions define the phase boundaries for different magnetic phases. Here, we designate Phases I, II and III as those corresponding to the lower-field, the 1/3 magnetization plateau regime, and higher-field metamagnetic transitions, respectively. Intriguingly, the phase diagram of GdTi$_3$Bi$_4$ shows a notable similarity to that of the well-known skyrmion-lattice compound Gd$_2$PdSi$_3$ [19], implying the possible existence of skyrmion phase in GdTi$_3$Bi$_4$. However, the Hall resistivity in GdTi$_3$Bi$_4$ is more complex, particularly in the region where $H_2 < H < H_3$.

To understand the origin of AHE in GdTi$_3$Bi$_4$, we conducted systematic magnetic force microscopy (MFM) measurements under varying magnetic fields. As displayed in Fig. 3a, the 4 K MFM data shows homogeneous magnetic signal at 5 T (the saturation field), with no observable domain walls. However, as the magnetic field decreases to 3.46 T, approaching the higher-field metamagnetic transition, elongated magnetic domains start to nucleate. As the field further deceases, more domains nucleate and keep extending in certain direction, finally these domains merge into each other to form periodic stripe phase. At the metamagnetic transition field around 3.40 T, the density of stripe domains maximizes, indicating a peak in both the number of domains and domain walls per unit area. Subsequently, the density of stripe domains reduces as the magnetic field moves away from the transition. Within the 1/3 plateau regime (Phase II in Fig. 2f), the MFM signal becomes homogeneous once again. As the magnetic field decreases to the second metamagnetic transition at a lower field, the stripe domain walls reappear, and they disappear as the field reduces to zero field subsequently. Compared to the higher-field metamagnetic transition, these stripe domains at the second metamagnetic transition align in a different orientation, while show similar periodicity. When a negative magnetic field is applied from 0 to -5 T, the domain walls exhibit the same field dependence (Supplementary Fig. 5).

By plotting the field dependence of the Root Mean Square (RMS) value of MFM images together with Hall and magnetization data, as shown in Figs. 3d–3f, it is clear that the field evolution of the stripe domain walls aligns well with the abrupt changes observed in Hall resistivity and magnetization.

The field evolution of the domain walls has also been confirmed at different temperatures (also see Supplementary Note 5). As the temperature increases, the low-field stripe phase is more robust against thermal fluctuation, as the orientation and shape of the stripe domains almost stay the same. The high-field stripe domains, however, seem to be more disconnected and disordered at higher temperatures. From the Fast Fourier Transform (FFT) analysis, above 8 K, a new stripe order forms, indicating a transition from single-Q stripe phase to two-Q phase. It is important to note that within the magnetization plateau regime and the range $H_2 < H < H_3$, there are no domain walls, ruling out their contribution to the AHE in these states and pointing to another possible origin, such as a skyrmion phase arising from chiral magnetic textures.

Recently, a domain wall skew scattering mechanism has been theoretically proposed and experimentally demonstrated in an A-type antiferromagnetic Weyl semimetal [52]. The contribution of domain walls to the Hall signal is expected to be proportional to the net magnetization at the domain walls. To investigate these contributions in GdTi$_3$Bi$_4$, the sample size was reduced to a few microns with a thickness of ~ 200 nm (Supplementary Fig. 6), considering that the domain wall size is on the micron scale (Fig. 3a). Figures 3b and 3c display the Hall resistivity and MR of a flake sample (the inset to Fig. 3b), respectively. In sharp contrast to the bulk sample, the Hall resistivity of the flake sample shows spikes rather than jumps at the metamagnetic transition fields, while the MR profiles remain similar (Fig. 3c). For comparison, the Hall resistivity at 4 K is also plotted in Fig. 3e. At the higher-field metamagnetic transition, the spike in Hall resistivity aligns well with the RMS signal in MFM measurements and the jump in the Hall resistivity of the bulk sample. However, the spike at the lower-field metamagnetic transition occurs at a lower field. This difference may be due to variations in domain wall skew scattering or magnetization between the flake and bulk samples. Therefore, unlike the bulk sample, the flake sample in GdTi$_3$Bi$_4$ exhibits both skyrmion-like AHE and an AHE enhanced by domain-wall skew scattering.

**Reverse magnetic chirality at metamagnetic transitions**

The behavior of stripe domain walls entering and exiting the magnetization plateau is highly intriguing and warrants further investigation. To understand the origin of the stripe domain walls and anomalous transverse transport in GdTi$_3$Bi$_4$, we consider the spin model on the Gd-based lattice in an external magnetic field defined by the following Hamiltonian:

$$H = J_1 \sum_{\langle i,j \rangle} \mathbf{S}_i \cdot \mathbf{S}_j + J_2 \sum_{\langle\langle i,j \rangle\rangle} \mathbf{S}_i \cdot \mathbf{S}_j + J_3 \sum_{\langle\langle\langle i,j \rangle\rangle\rangle} \mathbf{S}_i \cdot \mathbf{S}_j + K \sum_i (S_i^z)^2 + \sum_i \mathbf{B} \cdot \mathbf{S}_i$$

where $\langle i,j \rangle$, $\langle\langle i,j \rangle\rangle$, and $\langle\langle\langle i,j \rangle\rangle\rangle$ denote the summation over the nearest, next-nearest, and next-next-nearest exchange couplings. $J_1$, $J_2$, $J_3$, and $K$ stand for the Heisenberg exchange, and single ion anisotropy, respectively. Last, the parameter $B = g\mu_B H$ denotes the standard Zeeman's term, $\mu_B$ is the Bohr magneton, $g$ is the gyromagnetic factor of magnetic ions and $H$ is an external field. As shown in Fig. 4a, the $J_1$ and $J_3$ Heisenberg exchange coupling form into the one-dimensional zigzag chain along the $a$ axis, and the $J_2$ Heisenberg exchange coupling connects the different zigzag chain along the $b$ axis. Here, we adopt the parameters of $J_1 = J_3 = 0.2$ meV, $J_2 = -0.2$ meV, $K = 0.1$ meV, which means that the coupling of intra-chain is antiferromagnetic, while the coupling of inter-chain is ferromagnetic. Due to the high symmetries of space group *Fmmm* (No. 69), the nearest and next-nearest exchange coupling cannot allow the existence of Dzyaloshinskii–Moriya interaction, which can only be expected in the next-next-nearest exchange coupling.

Next, we perform parallel tempering Monte Carlo simulations to investigate the topological spin textures in GdTi$_3$Bi$_4$, utilizing the defined spin Hamiltonian. The lower panel in Fig. 4b displays the average magnetization per Gd site as a function of the magnetic field. Surprisingly, our simulation captures the main features observed in experiments (the upper panel in Fig. 4b). The magnetization stabilizes at approximately 2.2 $\mu_B$ over a range of magnetic fields from 1.4 T to 3.2 T. This stabilization results in a quasi-quantized 1/3 magnetization plateau, characterized by an exotic ferrimagnetic order where the spins are arranged in an up-up-down sequence. In this configuration, 16 out of 24 spins within the expanded 3×1×1 magnetic unit cell exhibit a magnetic moment of about 7 $\mu_B$ per Gd along the $c$ axis, as shown in the middle panel of Fig. 4c. It is noteworthy that the calculated magnetization plateau is slightly below the 1/3 level, a deviation attributed to the in-plane spin components. Furthermore, the out-of-plane magnetization shows similar stripe domain walls entering and exiting magnetization plateau, whereas the in-plane magnetization exhibits distinct shapes in these regions. As illustrated in Fig. 4c, the chirality of the two magnetic structures is +2 and -1 for Phase I and Phase III, respectively. At high magnetic fields, the magnetization reaches full saturation, aligning all magnetic dipoles completely. To summarize, our spin model demonstrates that the magnetic textures at two successive metamagnetic transitions possess reverse chirality, resulting in stripe domain walls.

As previously mentioned, many magnetic materials have been reported to exhibit the THE (referred to as the nominal AHE in this work) [13–29]. For typical chiral magnets, such as FeGe with helimagnetic order [17], MnSi with a conical structure [14,15], and Cr$_{1/3}$NbS$_2$ and CrNb$_3$S$_6$ with soliton lattices [53,54], the DMI has been identified as a crucial mechanism in stabilizing the magnetic skyrmions. Besides DMI, other mechanisms such as long-range magnetic dipolar interactions [55,56], frustrated exchange interactions (with and without geometric frustration) within centrosymmetric crystal [19–29,57] could also give rise to the THE. These factors are expected to favor a multiple-Q skyrmion lattice state in crystal lattice systems with high symmetry, such as hexagonal or tetragonal structures [57]. Among these highly symmetric systems, the magnetization of the centrosymmetric tetragonal magnet $GdRu_2Si_2$ somewhat resembles that observed in $GdTi_3Bi_4$ [57]. Within Phase II of the phase diagram for $GdRu_2Si_2$ (corresponding to Phases I, II, and III in $GdTi_3Bi_4$), a broad peak in Hall resistivity associated with nanometric square skyrmion lattices has been observed [57]. A similar scenario might occur in $GdTi_3Bi_4$, but detecting such nanoscale skyrmion lattices is beyond the resolution of our MFM setup. Moreover, the AHE in $GdTi_3Bi_4$ is more complex. For instance, in Phase III of $GdRu_2Si_2$ (corresponding to the range $H_2 < H < H_3$) in $GdTi_3Bi_4$, no THE is detected in $GdRu_2Si_2$, whereas it is present in $GdTi_3Bi_4$. Additionally, as mentioned earlier, domain walls also contribute to the AHE in $GdTi_3Bi_4$, further distinguishing it and placing it in a unique position. We speculate that the distinct AHE of $GdTi_3Bi_4$ arise from its unique magnetic structure, crystal structure, and symmetries—specifically, the zigzag chains imposed by two-fold symmetries originating from the distorted kagome lattice. This underscores the significance of the interplay between magnetism and crystal symmetries in enhancing the useful physical properties of materials.

## Conclusions

In this work, we systematically investigate the electronic band structure, and transport properties of the newly discovered magnetic kagome metal $GdTi_3Bi_4$. ARPES measurements demonstrates the presence of double FBs and multiple vHSs, and also uncovers band splittings in both magnetic and paramagnetic states, which is in sharp contrast to other $Ln Ti_3Bi_4$ siblings. Transport measurements on both bulk and flake sample reveal a unique AHE associated with a possible skyrmion phase. At the two successive metamagnetic transitions, stripe domain walls are revealed, potentially enhancing the AHE through a domain-wall skew scattering mechanism. Interestingly, our theoretical simulations demonstrate that the magnetic textures at these transitions exhibit reverse chirality, explaining the field evolution of AHE and domain walls in $GdTi_3Bi_4$. Compared with other skyrmion-host magnets, we propose that the synergy of magnetism and the quasi-1D zigzag chains imposed by crystal symmetries within kagome lattice plays a crucial role in $GdTi_3Bi_4$. This provides a unique strategy to enhance the versatility and tunability for material design, opening intriguing avenues for future spintronic applications.

**Methods**

**Sample preparation**

GdTi$_3$Bi$_4$ single crystals were grown using a self-flux method with an elemental ratio of Gd:Ti:Bi of 1.2:3:20. Gd (99.95%), Ti (99.99%), and Bi (99.999%) were cut into small pieces, mixed, and placed in an alumina crucible. The crucible was then sealed in a quartz tube under partial argon pressure. The sealed tube was heated to 1,050 °C over 12 hours and maintained at that temperature for 24 hours. It was then slowly cooled to 500 °C at a rate of 2 °C per hour. Single crystals were obtained by removing the flux through centrifugation. Compositional analysis of the as-grown single crystals of GdTi$_3$Bi$_4$ is shown in Supplementary Fig. 1.

**Electrical and thermodynamic measurements**

For transport measurements, a single crystal was cut into a bar shape. A standard six-probe method was used for the longitudinal resistivity and transverse Hall measurements. Electrical transport data were collected in a physical property measurement system (PPMS, Quantum Design). Magnetic susceptibility and specific heat measurements were performed in a magnetic property measurement system (MPMS, Quantum Design) and a PPMS, respectively.

**ARPES measurements**

High-resolution ARPES measurements were performed using the CASSIOPEE beamline at the SOLEIL synchrotron and the "Dreamline" beamline at the Shanghai Synchrotron Radiation Facility. The energy and angular resolutions were set to better than 10 meV and 0.1°, respectively. Samples were cleaved *in situ*, yielding flat and mirrorlike (001) surfaces. During the experiments, the sample temperature was kept at ~9 K or ~20–25 K if not specified otherwise, and the vacuum conditions were maintained better than 2×10$^{-10}$ Torr. We used linearly horizontal polarized photons for all measurements.

**MFM measurements**

The GdTi$_3$Bi$_4$ single crystals were exfoliated onto Si substrates using Scotch tapes and the fresh exfoliated surface was capped by 10nm thick Au film for surface protection and electrical potential balance. The cryogenic MFM experiments were performed on a commercial atomic force microscope (attoAFM I) using commercial cantilevers (spring constant $k \approx 3.8 \text{ N/m}$ and resonance frequency $\approx 85$ kHz) based on a closed-cycle helium cryostat (attoDRY2100). An out-of-plane magnetic field was applied via a superconducting magnet. MFM images were carried out in constant height mode with lift height around 200 nm. The MFM signals, the change of cantilever resonance frequency, is proportional to the out-of-plane stray field gradient. Electrostatic interaction was minimized by balancing the tip-surface potential difference. The blue regions in the MFM images represent fully polarized ferromagnetic regions where the magnetization is parallel with the external field, which corresponds to the high-field magnetization plateau. The cyan regions represent up-up-down ferrimagnetic domains, corresponding to the 1/3 magnetization plateau. The red regions represent antiferromagnetic ground state, which corresponds to zero-field magnetization plateau on the *M* (*H*) curves.

**DFT calculation**

Our calculations are performed using the projector augmented wave method [58,59] as implemented in the Vienna *ab initio* Simulation package [60,61]. Experimental lattice structure is adopted in our calculations. The exchange-correlation functional is treated within a generalized gradient approximation parametrized by Perdew, Burke, and Ernzerhof [62]. In the self-consistent calculations, the cutoff energy for the plane-wave expansion is 500 eV, and the *k*-point sampling grid of the Brillouin zone is 5×6×7. To simulate the paramagnetic state, the 4*f* electrons of Gd are treated used as core electrons. For the calculation of ferromagnetic state, the local moments of Gd atoms are aligned in the *z* direction.

**Atomistic spin dynamics simulations**

Here, we employ parallel tempering Monte Carlo simulations using the Sunny package to determine the energy-minimized spin textures at a fixed temperature of 2 K [63]. The spin textures are derived from a 20 × 20 × 20 supercell, with $1 \times 10^4$ Monte Carlo steps executed for each magnetic field increment, ranging from 0 T up to 7 T.

**Fabrication and characterization of flake sample**

The flake sample is exfoliated from single crystal GdTi$_3$Bi$_4$ using conventional scotch tapes and is transferred to a Si substrate coated with 300 nm SiO$_2$. Contact metal is deposited immediately after a stencil mask aligned to the flake. All fabrication processes are carried out in a glove box with oxygen and water lower than 0.1 ppm to prevent the degradation. The electrical transport is performed in a PPMS. Two lock-in amplifiers SR 830 are used to measure $R_{xx}$ and $R_{xy}$ simultaneously at 17 Hz.

## Data availability

The data that support the findings of this study are available from the corresponding authors upon request.

generalized spins as SU (N) coherent states. *Phys. Rev. B* **106**, 235154 (2022).

## Acknowledgments

This work was financially supported by the Deutsche Forschungsgemeinschaft (DFG) under SFB1143 (Project No. 247310070), the Würzburg-Dresden Cluster of Excellence on Complexity and Topology in Quantum Matter—ct.qmat (EXC 2147, Project No. 390858490), and Grant No. QUASTFOR5249-449872909. E.J.C. and N.M. acknowledge the financial support from the Alexander von Humboldt Foundation. Y. Z. acknowledges the support by the National Science Foundation Materials Research Science and Engineering Center program through the UT Knoxville Center for Advanced Materials and Manufacturing (DMR-2309083). E.J.C. thanks Wei Xia and Yang Xu for the helpful discussions on the analysis of transport data.

## Author Contributions

E.J.C. convinced the idea and designed the experiments. E.J.C. grew the single crystals, and conducted magnetization and electrical transport measurements of bulk sample. E.J.C. analyzed the electrical transport data. R.L., A.F., F.B., P.F.D., O.S., S.C., and S.C.W. conducted ARPES measurements. R.L. analyzed the ARPES data and drafted the accompanying discussions. X.T.Y. and W.B.W. performed MFM measurements and analysis. N.M., S.Z., and Y.Z. conducted Monte Carlo (MC) simulations. S.M.N. performed the DFT calculations for the band structure in PM and FM states. B.Q.S. and T.P.Y. fabricated the flake sample and performed the measurements. R.K. and W.S. helped to measure heat capacity. U.B. performed the compositional analysis. E.J.C., Y.Z., R.L., W.B.W. and C.F. supervised the project. E.J.C., N.M., X.T.Y., B.Q.S., and R.L. contributed equally to this work. E.J.C. wrote the paper with input from all coauthors.

## Competing interests

The authors declare no competing interests.

## Additional Information

**Supplementary information** is available for this paper at the URL inserted when published.
**Correspondence** and requests for materials should be addressed to E.J.C. (Erjian.Cheng@cpfs.mpg.de), R.L. (lourui09@gmail.com), Y.Z. (yangzhang@utk.edu), W.B.W.

(wangwb1@shanghaitech.edu.cn) or C.F. (Claudia.Felser@cpfs.mpg.de).

**Figure 1**

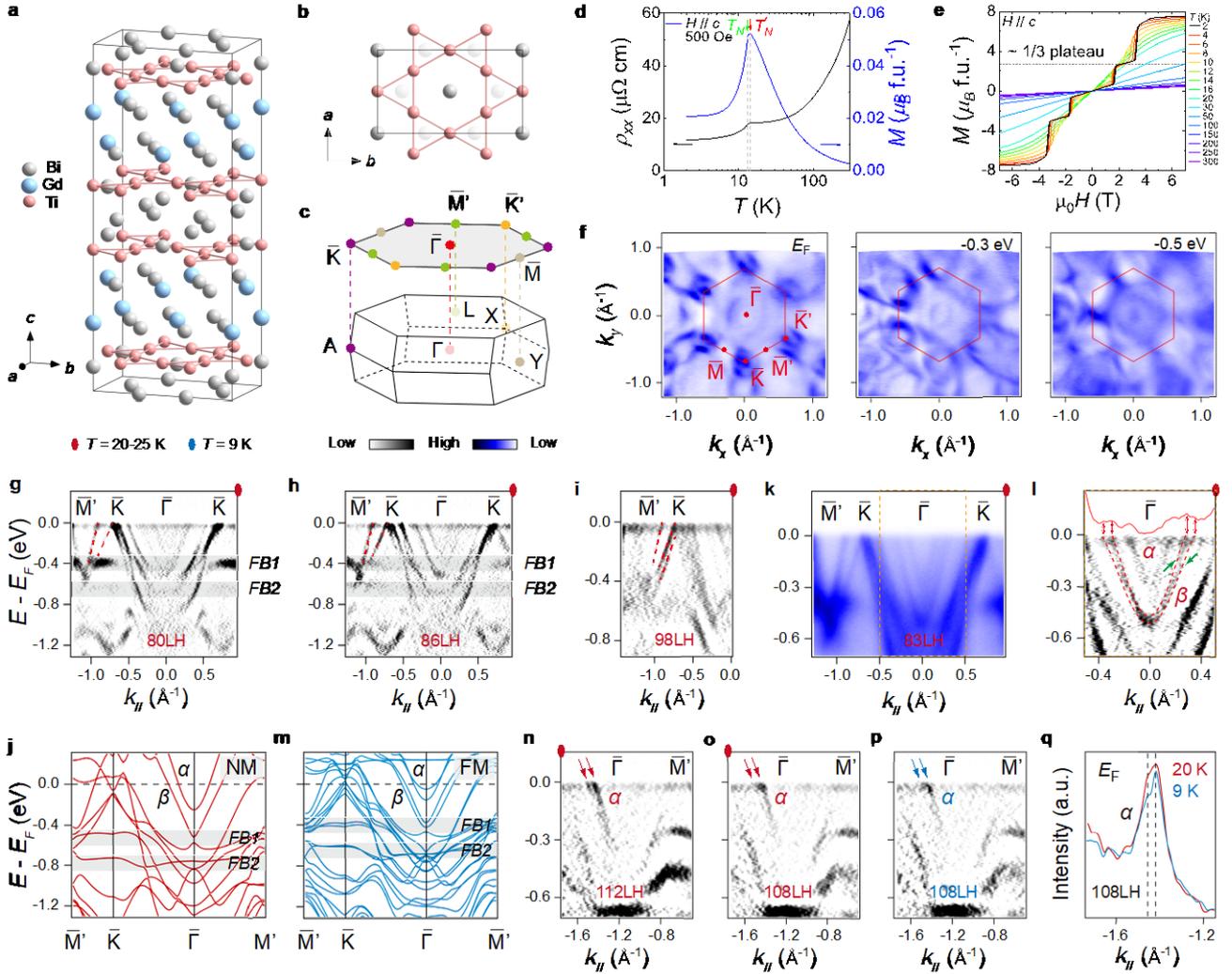

**Figure 1 | Crystal structure and electronic structure of GdTi$_3$Bi$_4$.** Side view (**a**) and bottom view (**b**) of the GdTi$_3$Bi$_4$ crystal structure, respectively. (**c**) Bulk and pseudo-hexagonal projected BZs with the high-symmetry points. (**d**) Temperature dependence of the zero-field resistivity $\rho_{xx}$ (black curve) and magnetization (blue curve) with field applied along the $c$ axis. Red ($T'_N$) and green arrows represent the temperatures where two consecutive magnetic transitions take place. (**e**) Magnetic field dependence of magnetization at various temperatures with field applied along the $c$ axis. A nearly one-third of the saturated magnetization (the 1/3 plateau) has been identified. (**f**) Constant-energy ARPES intensity plots ($h\nu = 86$ eV) at energies of 0, -0.3, and -0.5 eV, respectively. The red solid lines indicate the (001)-projected BZs. (**g-i**) Second derivative intensity plots ($T = 25$ K) recorded along the $\bar{\Gamma}$-$\bar{K}$-$\bar{M}'$ direction with 80- (**g**), 86- (**h**), and 98-eV (**i**) photons, respectively. The red dashed curves are guides to the eye for the near-$E_F$ band dispersions around $\bar{M}'$ point. (**j**) DFT calculated bulk band structure

along the $\overline{M}'$-$\overline{K}$-$\overline{\Gamma}$-$\overline{M}'$ lines ($k_z = \pi$ plane) in the paramagnetic phase of GdTi$_3$Bi$_4$. (**k**) ARPES intensity plot ($T$ = 25 K, $hv$ = 83 eV) measured along the $\overline{\Gamma}$-$\overline{K}$-$\overline{M}'$ direction. (**l**) Second derivative intensity plot of (**k**) over the momentum range of -0.5 < $k_{//}$ < 0.5 Å$^{-1}$, as indicated by the orange dashed rectangle in (**k**). The red dashed curves are guides to the eye for the split $\beta$ band. Due to matrix element effects, some portions of the split bands are not visible in the right branch, it thus looks like a "kink" feature as guided by the green arrows. Inset: MDC taken at $E_F$ with the red arrows showing the $k_F$ crossings of split states. (**m**) Same as (**j**) but with a ferromagnetic order on the Gd sites. (**n,o**) Second derivative intensity plots ($T$ = 20 K) recorded along the $\overline{\Gamma}$-$\overline{M}'$ direction with 112- (**n**) and 108-eV (**o**) photons, respectively. (**p**) Same as (**o**) but at $T$ = 9 K. The red and blue arrows in (**n-p**) indicate the splitting of $\alpha$ band. (**q**) Temperature-dependent MDCs taken around $E_F$ with the dashed lines showing the $k_F$ crossings of split $\alpha$ band.

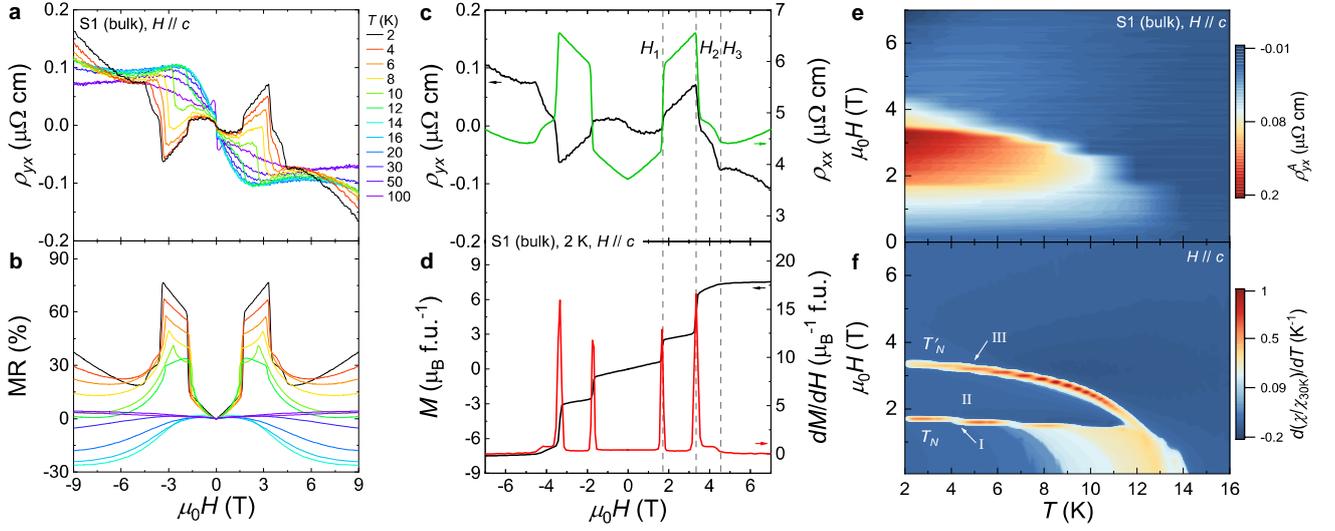

**Figure 2 | Anomalous transverse transport measurements of GdTi$_3$Bi$_4$.** (**a**) Transverse transport ($\rho_{yx}$) (bulk sample, S1) with current $j$ and magnetic field applied along the $a$ and $c$ axis, respectively. (**b**) Magnetoresistance (MR) at different temperatures, which is defined as $\mathrm{MR} \equiv 100 \times [\rho_{xx}(H) - \rho_{xx}(0)]/\rho_{xx}(0)$. (**c,d**) A comparison of Hall resistivity, longitudinal resistivity, magnetization, and the first derivative of magnetic susceptibility at 2 K with magnetic field applied along the $c$ axis. (**e**) Contour plot of the anomalous Hall resistivity ($\rho_{yx}^{A}$). The background color represents the values of $\rho_{yx}^{A}$. (**f**) Contour plot of magnetization as a function of the magnetic field applied along the $c$ axis during the zero-field cooling process. The background color represents the differential values of magnetization with respect to the magnetic field. Phases I and III denote the states at the lower- and higher-field metamagnetic transitions, respectively. Phase II specifically corresponds to the regime of the 1/3 magnetization plateau.

**Figure 3**

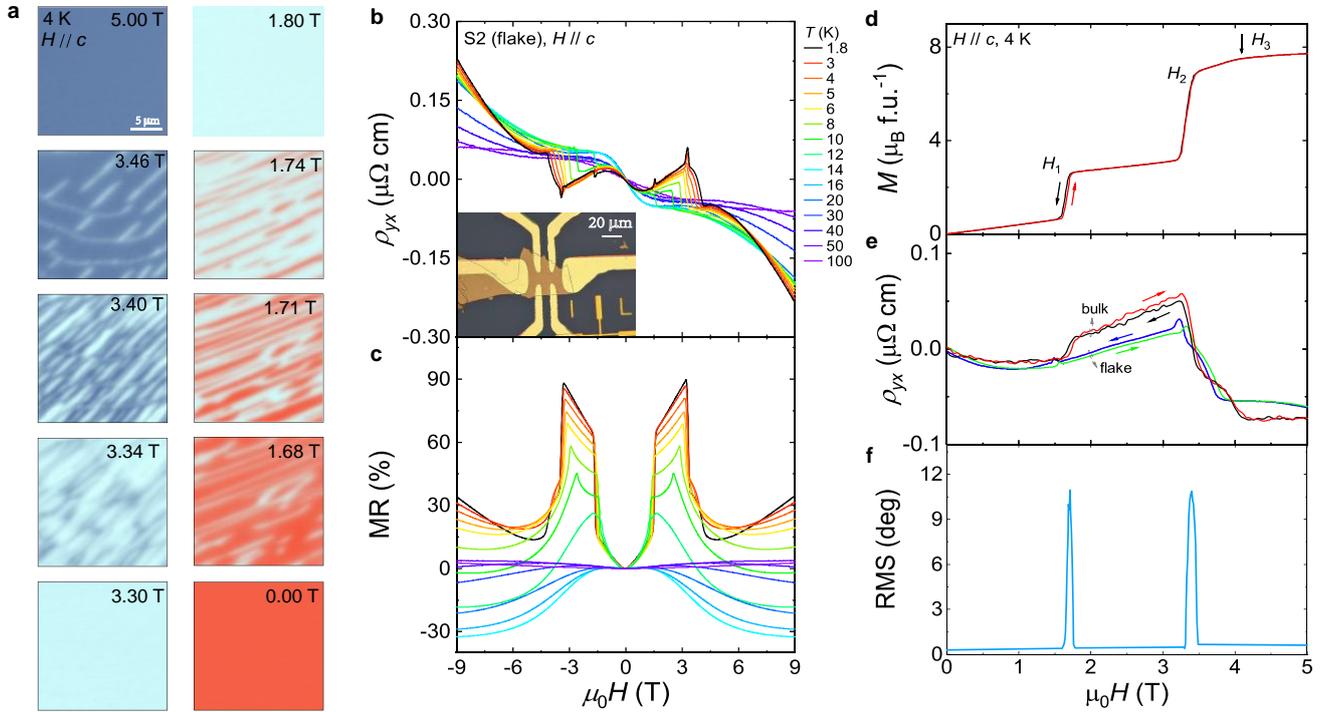

**Figure 3 | Magnetic force microscopy (MFM) topography of the domain walls with the evolution of magnetic fields in GdTi$_3$Bi$_4$.** (**a**) MFM images taken at 4 K with magnetic field applied parallel to the *c* axis. Stripe-like domain walls exist at two metamagnetic transitions. (**b**) Transverse transport ($\rho_{yx}$) (flake sample, S2) with magnetic field applied along the *c* axis. (**c**) The corresponding magnetoresistance (MR) at different temperatures. A comparison of magnetization (**d**) and Hall resistivity, and MFM signal at 4 K with magnetic field applied along the *c* axis. The arrows represent the field-sweeping direction.

**Figure 4**

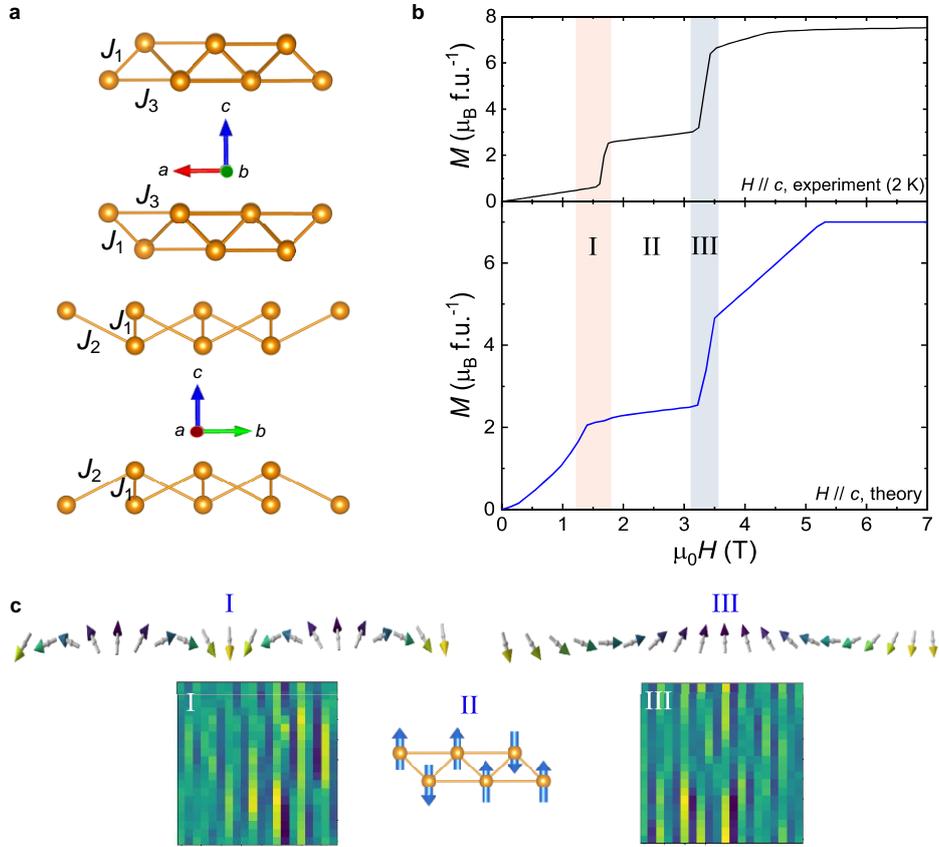

**Figure 4 | Chiral magnetic structure of GdTi$_3$Bi$_4$ in magnetic fields.** (**a**) Schematic illustration of the nearest, next-nearest, and next-next-nearest exchange couplings. (**b**) A comparison of the magnetization between experiments (2 K, the upper panel) and calculations (the lower panel). Magnetic field is applied parallel to the *c* axis. The pink (I) and blue (III) color bars represent the regime of two metamagnetic transitions at ~1.7 T and ~3.3 T, respectively. II represents the regime of 1/3 magnetization plateau. (**c**) Both in-plane and out-of-plane magnetic configurations before and after the plateau; arrows indicate the in-plane components, while domain walls represent the out-of-plane components. The out-of-plane magnetic configuration under the plateau has also been illustrated (II).